\documentclass[aps,prl,twocolumn,superscriptaddress,floatfix]{revtex4-2}
\usepackage[english]{babel}
\usepackage{amsmath,amssymb,amsfonts}
\usepackage{graphicx}          
\usepackage{bm}                
\usepackage{xcolor}
\usepackage{siunitx}           
\usepackage{hyperref}          
\usepackage{cleveref}          
\usepackage{booktabs}          
\usepackage{multirow}          
\usepackage{dcolumn}           
\usepackage{physics}           
\usepackage{upgreek}           
\usepackage{subfigure}         
\usepackage{natbib}            

\hypersetup{
  colorlinks = true,
  citecolor  = blue,
  linkcolor  = blue,
  urlcolor   = blue
}

\begin{document}

\title{Unlocking the O-Band Coherent Transmission: high-power, broadband soliton microcomb}

\author{Dmitrii Stoliarov$^{\dagger}$}
\email{d.stoliarov@aston.ac.uk} 
\affiliation{Aston Institute of Photonic Technologies, Aston University, B4 7ET Birmingham, UK}

\author{Nikolay G. Pavlov$^{\dagger}$}
\affiliation{Enlightra, Rue de Lausanne 64, Renens, 1020, VD, Switzerland}

\author{Aleksandr Donodin}
\affiliation{Aston Institute of Photonic Technologies, Aston University, B4 7ET Birmingham, UK}

\author{Daniel J. Elson}
\affiliation{KDDI Research, 2-1-15 Ohara, Fujimino, Saitama, 356-8502, Japan}

\author{Vitaly Mikhailov}
\author{Jiawei~Luo}
\affiliation{Lightera Laboratories, 19 Schoolhouse Rd., Somerset, New Jersey 08873, USA}

\author{Sergey Koptyaev}
\affiliation{Enlightra, Rue de Lausanne 64, Renens, 1020, VD, Switzerland}

\author{Robert Emmerich}
\affiliation{Fraunhofer Heinrich-Hertz-Institute, Einsteinufer 37, 10587 Berlin, Germany}

\author{Ruben S. Luis}
\author{Hideaki Furukawa}
\affiliation{The National Institute of Information and Communications Technology, Tokyo 184-8795, Japan}

\author{Colja Schubert}
\author{Ronald Freund}
\affiliation{Fraunhofer Heinrich-Hertz-Institute, Einsteinufer 37, 10587 Berlin, Germany}

\author{Yuta~Wakayama}
\author{Takehiro Tsuritani}
\affiliation{KDDI Research, 2-1-15 Ohara, Fujimino, Saitama, 356-8502, Japan}

\author{David J. DiGiovanni}
\affiliation{Lightera Laboratories, 19 Schoolhouse Rd., Somerset, New Jersey 08873, USA}

\author{John D. Jost}
\author{Maxim Karpov}
\affiliation{Enlightra, Rue de Lausanne 64, Renens, 1020, VD, Switzerland}

\author{Sergei~K.~Turitsyn}
\affiliation{Aston Institute of Photonic Technologies, Aston University, B4 7ET Birmingham, UK}


\date{\today} 

\begin{abstract}
The O-band (1260–1360 nm), located near the minimum chromatic dispersion of standard single-mode fiber, is an important transmission window for short-reach data-center interconnects. Its broader use is limited by the availability of scalable multi-wavelength, high-power and low-noise light sources. Here we demonstrate a high-power O-band soliton microcomb architecture combining an 834~GHz self-injection-locked (SIL) $\mathrm{Si_3N_4}$ microcomb spanning 1050--1650~nm with a single-stage bismuth-doped phosphosilicate fiber amplifier. The system delivers $>0$~dBm per carrier for 21 O-band comb lines over a 100~nm bandwidth, exhibiting low-noise operation and a 5~dB variation in amplifier gain. We validate the amplified source in a 25-km coherent WDM transmission experiment using 32-GBd dual-polarisation 64-QAM. Comb carriers spanning the O-band are characterised under simultaneous WDM loading, yielding an aggregate GMI-estimated throughput of ~7.23 Tb/s and a post-FEC throughput of ~6.94 Tb/s. 
This approach establishes a practical route towards broadband, high-power O-band microcomb sources for multi-terabit optical interconnects.
\end{abstract}

\maketitle


\begingroup
\renewcommand\thefootnote{}
\footnotetext{$^\dagger$These authors contributed equally.}
\endgroup


The rapid expansion of artificial intelligence (AI) workloads and cloud services is driving unprecedented demand for communication capacity within data centers, where short-reach optical interconnects form the backbone of distributed computing infrastructure. Capacity scaling in these systems increasingly relies on wavelength-division multiplexing (WDM), requiring compact multi-wavelength light sources capable of delivering many stable carriers while maintaining strict constraints on power per line, footprint, and yield \cite{cheng2018recent,xie2022scaling,Zhu2026NextGen}. The O-band (1260–1360 nm) is particularly attractive for intra- and inter-data-center links because of its low chromatic dispersion, reducing the need for dispersion compensation. This spectral window provides $\sim$ 17.6 THz of usable bandwidth, offering significant potential for capacity scaling. 
However,  future network capacity scaling will require increasing the number of wavelength channels and, where appropriate, adopting coherent and co-packaged architectures \cite{buscaino2021external}. Developing scalable multi-wavelength O-band light engines therefore remains a key challenge for next-generation data-center architectures.
Several approaches have been explored to address this challenge. Arrays of semiconductor lasers and multi-wavelength diode bars can provide multiple carriers but scale unfavourably as channel counts increase, due to fabrication tolerances, thermal management, and control complexity \cite{hong2025versatile,zheng2013efficient,lu2015multi,shi2014high}. Electrically driven quantum-dot (QD) mode-locked lasers offer a compact comb-based alternative in the O-band \cite{NarayananVenkatasubramani:25,pan2020quantum}, yet their spectral flexibility and channel scalability remain constrained by the gain bandwidth (10-12  nm) of the semiconductor medium and relatively small line spacing of tens of gigahertz \cite{Bernal2024NatCommun, buyalo2024efficient}.

\begin{figure*}[!t]%
\centering
\includegraphics[width=0.9\textwidth]{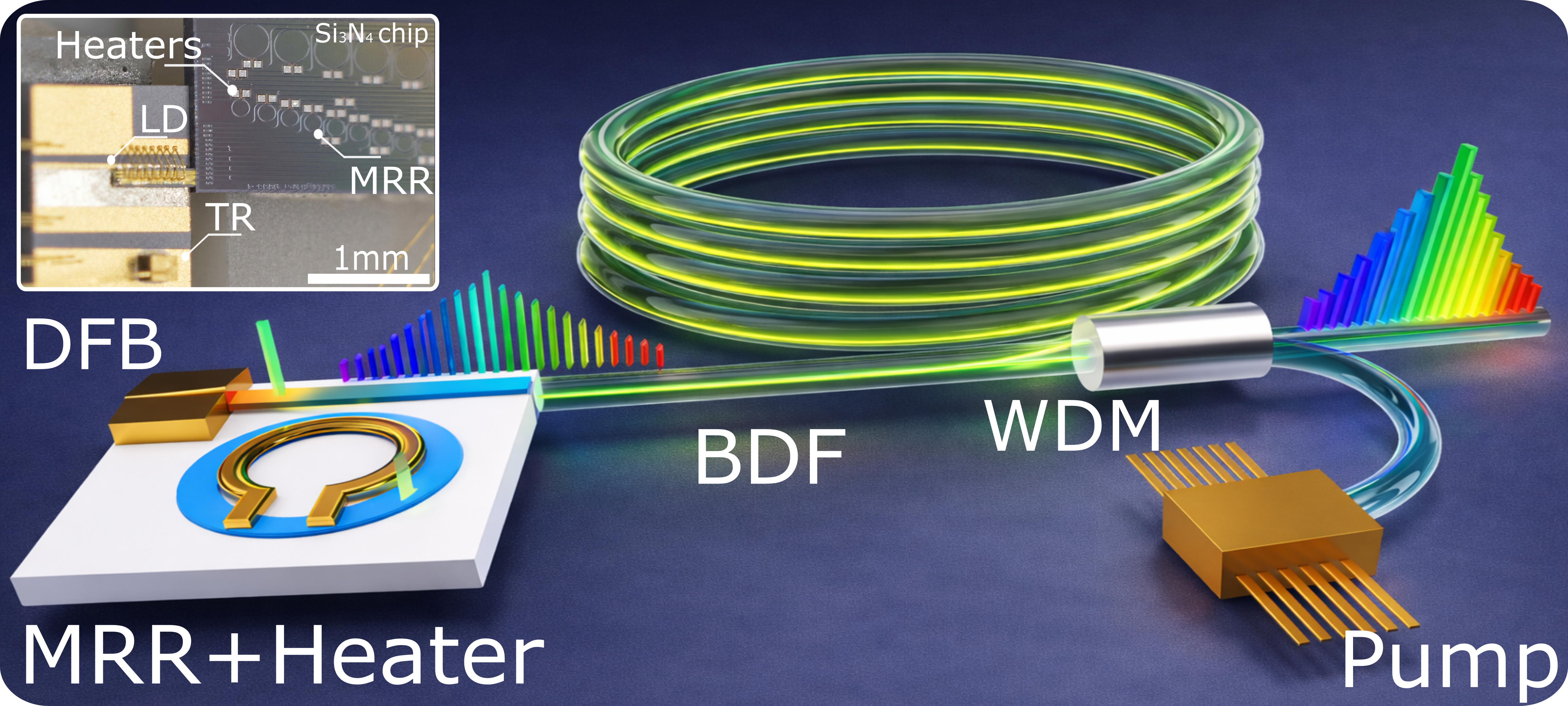}
\caption{\textbf{High-power O-band microcomb source illustration.} Schematic of the self-injection-locked (SIL) microcomb and bismuth-doped fiber (BDF) amplifier power-scaling chain. A distributed-feedback (DFB) laser diode (LD) is self-injection-locked to a Si\textsubscript{3}N\textsubscript{4}  microresonator (MMR) with integrated heaters for thermal tuning of the resonance frequencies. The comb output is amplified in a BDF pumped at 1150~nm via a wavelength-division multiplexing (WDM) coupler. Coloured spectral components are illustrative (shown in the visible for clarity rather than at near-infrared wavelengths) and indicate amplification of the initial comb after the BDFA stage. Inset: photograph of the  SIL package. LD: DFB laser butt-coupled to the Si\textsubscript{3}N\textsubscript{4} chip and the temperature-control assembly (TR-thermistor and heaters).}

\label{fig:fig1}
\end{figure*}
Kerr microcombs based on high--Q silicon nitride ($\text{Si}_3\text{N}_4$) microresonators offer a fundamentally different approach, providing mutually coherent, precisely spaced carriers with hundreds of gigahertz spacing suitable for wider WDM grids such as Local Area Network Wavelength Division Multiplexing (LAN-WDM). Their small footprint and scalable CMOS process fabrication offer an alternative to multi-wavelength DFB (Distributed Feedback) laser arrays or QD combs for data-center interconnects \cite{corcoran2025optical}. 
Microcombs have facilitated significant advancements in optical communications, including massively parallel intensity-modulated direct detection (IM/DD), coherent links, and high-capacity silicon-photonic systems.\cite{pirmoradi2025integrated,rizzo2023massively,shu2022microcomb, Riemensberger2020,feldmann2021tensorcore,bai2023ppu}. In the C-band, compact self-injection-locked (SIL) architectures have recently enabled electrically pumped soliton microcombs without the need for external narrow-linewidth lasers \cite{stern2018battery,raja2019electrically,shen_integrated_2020,xiang2021laser,Ulanov2024}. In this regime, Rayleigh back-scattering from the microresonator stabilizes the pump laser frequency while simultaneously enabling coherent comb generation \cite{oraevsky2001frequency,kondratiev2017self,kondratiev2020numerical,Kondratiev2023}.

While low-noise microcombs are well-developed in the C-band \cite{Kippenberg2018}, their translation to the O-band has remained challenging and has only been demonstrated with external pumping using an amplified tunable laser \cite{marchand2021soliton}. In addition to higher optical losses at shorter wavelengths, practical deployment has been hindered by insufficient per--line comb power and the lack of broadband, low--noise amplification technologies capable of uniformly boosting multiple coherent carriers across the full O--band transmission window.

Semiconductor optical amplifiers (SOAs) offer compact integration but typically exhibit wavelength-dependent gain, four-wave mixing, nonlinear distortions, and excess amplified spontaneous emission when amplifying multi-line sources  \cite{rombouts2024design,liu2019high, lawniczuk2015four}. For amplification with a wideband microcomb, the narrow-band gain and out-of-band gain shift limit the application of O-band SOAs.
Rare-earth-doped fiber amplifiers such as praseodymium-doped systems provide O-band gain but rely on non-silica host materials and do not provide sufficiently flat amplification across the entire transmission window \cite{ mirza2021performance,oxenlowe2025optical}. 
One of the most promising alternatives is bismuth-doped fiber (BDF) with bismuth active centers formed in phosphosilicate glass, which offers substantially broader potential gain coverage \cite{bufetov2014bi}. Over the last decade, amplifiers based on phosphosilicate and germanium glass have demonstrated broad gain across the  O+E-bands (1260-1460 nm) \cite{ososkov2021pump,donodin2024band}. Phosphosilicate BDF-based amplifiers for single-frequency sources have already demonstrated low noise figures (NFs) \cite{donodin2023pump} and high output powers \cite{stoliarov2026high}. Their application to high-capacity coherent O-band transmission has also been demonstrated \cite{mikhailov2019simple, Donodin2026ECOC}, however, achieving high and relatively flat gain across the entire 1260–1360 nm O-band remains challenging. As a result, wideband O-band microcomb sources have remained power-limited and difficult to deploy in wavelength-division multiplexed systems.
Here we demonstrate a high-power 834~GHz O-band soliton microcomb platform that overcomes these limitations by combining a self-injection-locked $\text{Si}_3\text{N}_4$ microresonator comb with a broadband bismuth-doped phosphosilicate fiber amplifier (BDFA) designed for flat gain across the O-band. The resulting system produces a soliton microcomb spanning more than 600~nm (1050--1650~nm) and simultaneously amplifies 21 coherent carriers across the full O-band with powers exceeding 0~dBm per line, preserving the initial comb spectral envelope.  We further evaluate the performance of the amplified microcomb in a coherent 25-km data transmission experiment. Generated comb lines used as carrier data encoded using 32~GBd 64--quadrature amplitude modulation (64--QAM), resulting in an aggregate data rate of 6.94~Tb/s for twenty comb lines. The 32-GBd DP-64QAM transmission demonstrates that the amplified O-band microcomb supports high-order coherent modulation across a broad wavelength range, highlighting its potential for high-capacity optical communication systems. 


\begin{figure*}[!t]%
\centering
\includegraphics[width=0.8\textwidth]{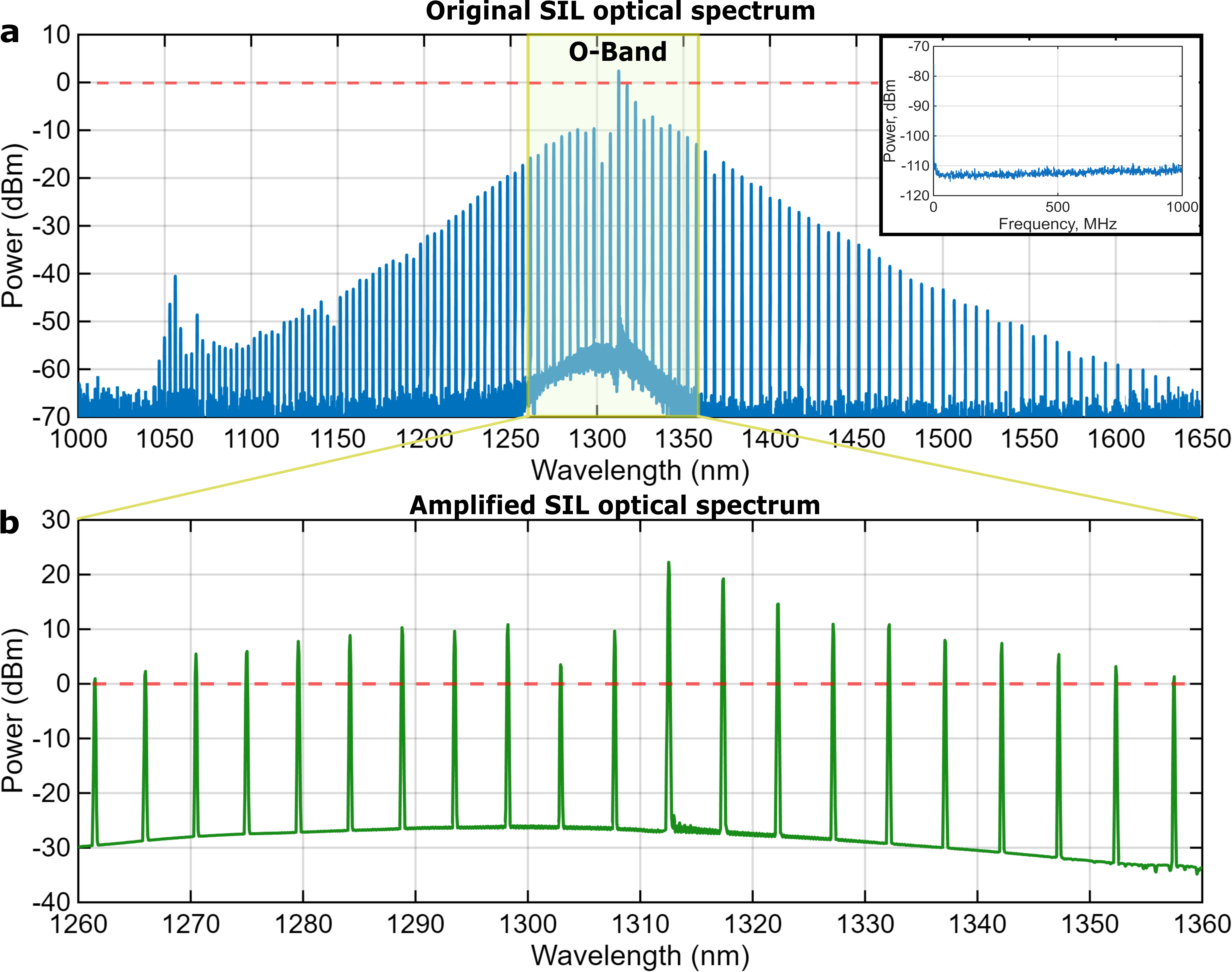}
\caption{\textbf{Optical spectrum of a SIL microcomb pumped at 1312~nm.} \textbf{a:} Optical spectrum of the SIL microcomb (FSR=834~GHz) before amplification. Inset: radio-frequency (RF) amplitude-noise spectrum measured on a photodiode shows low--noise state of the SIL microcomb. The red dashed line indicates the 0~dBm power level for reference.  \textbf{b:} optical spectrum of amplified SIL after the BDFA stage, showing 21 comb lines spanning 1260--1360~nm with powers exceeding 0~dBm per microcomb line (dashed red line is 0~dBm level for reference).}

\label{fig:fig2}
\end{figure*}

\begin{figure*}[!t]%
\centering
\includegraphics[width=1.\textwidth]{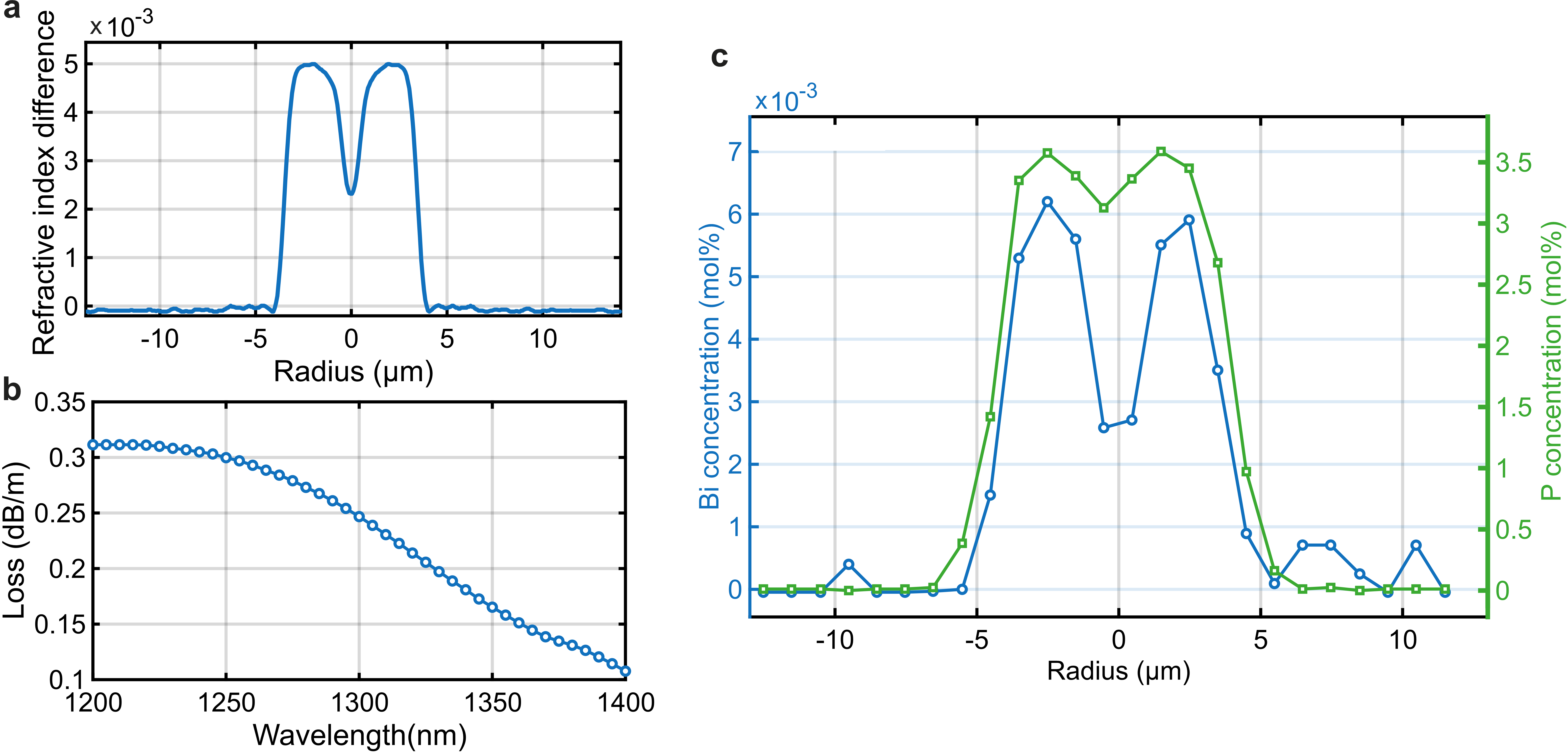}
\caption{\textbf{Bismuth doped fiber.} \textbf{a}: Designed BDF refractive index profile. \textbf{b}: Loss spectrum versus wavelength of the BDF. \textbf{c}:  the measured fiber \(\mathrm{Bi_2O_3}\) and \(\mathrm{P_2O_5}\) concentration profiles .} 

\label{fig:figSI2}
\end{figure*}

\section{RESULTS}
\textbf{Self-injection locking microcomb} 

Fig.\ref{fig:fig1} demonstrates a sketch of the amplified O--band microcomb. The inset shows a self-injection locking microcomb where DFB laser diode is pumping a $\text{Si}_3\text{N}_4$ ring resonator.
In the SIL regime, the microresonator plays a dual role: stabilizing the laser diode via back-reflected light from microresonator inhomogeneities, while simultaneously generating a low-noise microcomb. Note that the SIL configuration requires no optical isolators for the laser diode. Compared to a standalone DFB laser, the SIL regime provides high tolerance to external back-reflections \cite{Xiang2023}. Until now, most microcomb demonstrations have been conducted in the C-band, where $\text{Si}_3\text{N}_4$ exhibits lower losses and high-power lasers are more readily available \cite{Kippenberg2018}. Here, we utilize an O-band SIL comb based on a commercially available O-band DFB laser diode with a 1~mm chip length, a maximum output power of 180~mW, a maximum current of 500~mA and a center wavelength of 1312~nm.

\begin{figure*}[!t]%
\centering
\includegraphics[width=1.0\textwidth]{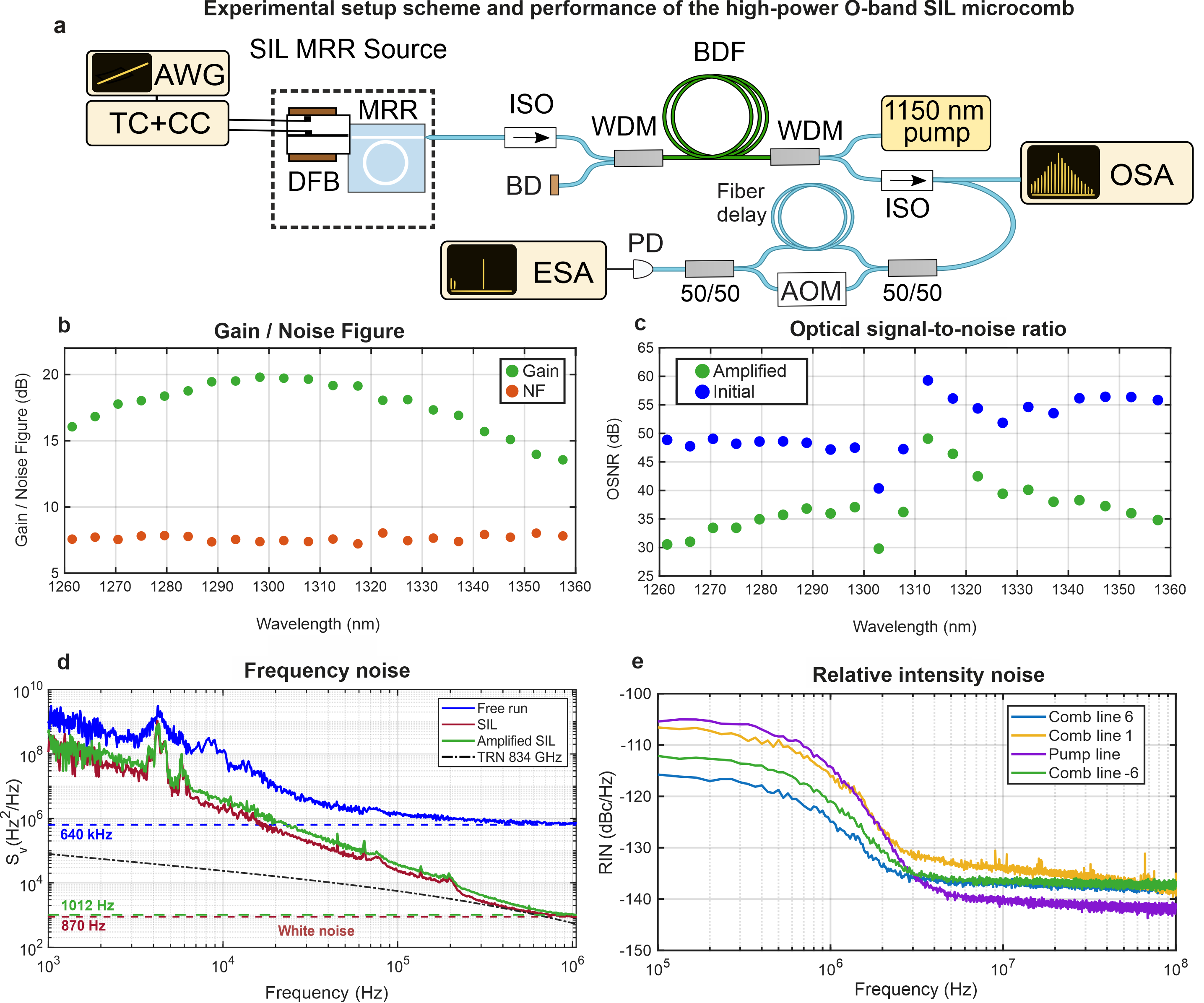}
\caption{\textbf{Experimental setup for high-power O-band SIL microcomb measurements.}  \textbf{a:} Experimental setup: AWG--arbitrary waveform generator; TC+CC--temperature and current controllers; LD--DFB laser diode, MMR-$\text{Si}_3\text{N}_4$--silicon nitride microresonator; ISO--fiber isolator; WDM--wavelength-division multiplexing coupler; BDF--bismuth-doped fiber; BD--Beam dump; OSA--optical spectrum analyzer; AOM--acousto-optic modulator; ESA--electrical spectrum analyzer. \textbf{b:}  BDFA gain spectrum (green dots) and corresponding noise figure (NF - orange dots) calculated from the amplified comb lines. \textbf{c:} Optical signal-to-noise ratio (OSNR) of the SIL comb before (blue dots) and after BDFA amplification (green dots). \textbf{d:} Frequency-noise for the free-running laser diode (blue curve), for laser diode in the SIL regime (red curve) and for BDFA amplified laser diode in the SIL regime (green curve), blue, red and green dashed curves are white noise levels for free--run LD, SIL LD and amplified SIL LD accordingly; dashed black curve is the calculated thermorefractive-noise (TRN) for the 834~GHz microresonator. \textbf{e:} Relative intensity noise (RIN) of selected amplified comb lines measured from 100 kHz to 100 MHz.} 
\label{fig:fig4}
\end{figure*}

The $\text{Si}_3\text{N}_4$ microresonator was fabricated at a commercial foundry using an 800~nm thick and $2\,\mu\text{m}$ wide waveguide geometry for the microresonator. The loaded Q--factor of the microresonator exceeds $1 \times 10^6$ in the O-band. The microresonator radius is $28\,\mu\text{m}$, corresponding to a FSR of 834~GHz. The geometry was designed to satisfy optimal coupling conditions for a 1310 nm pump wavelength and to provide the anomalous dispersion properties necessary for low-noise soliton state generation \cite{Kippenberg2016}. The 834~GHz microresonator FSR was selected to highlight the versatility of the approach, as it is close to the 800~GHz channel spacing used in LAN-WDM for short-reach high-speed networking \cite{li2022oband,andrianopoulos2024integrated}. For coherent transmission and IM/DD, the 800 GHz LAN-WDM grid is feasible, offering acceptable wavelength tolerances and good link margin. Based on this, LAN-WDM can utilize simpler optical components and avoid some of the stringent wavelength-control requirements found in dense WDM (DWDM) \cite{brackett1990}.

A high-power O-band LD is soldered onto an AlN submount with dimensions of $3 \times 2$ mm. The submount features a 10 k$\Omega$ NTC thermistor for laser diode temperature stabilization. The $\text{Si}_3\text{N}_4$ microresonator is butt-coupled to the LD on one side and coupled to a single-mode polarization-maintaining (PM) fiber pigtail on the opposite side (see the top-left inset of Fig.\ref{fig:fig1}). Finally, the $\text{Si}_3\text{N}_4$ ring is equipped with an integrated heater used for SIL frequency and phase control \cite{Joshi16}. The LD, $\text{Si}_3\text{N}_4$ chip, and output fiber are fully encapsulated in a standard 14-pin butterfly package ($30 \times 12.7 \times 8.7$~mm), which features an internal thermoelectric cooler (TEC) element.

The butterfly-packaged system exhibits turn-key operation. The DFB LD was driven by a Thorlabs current and TEC controller (ITC 4001). The laser diode frequency was tuned via current to match a microresonator resonance mode. Only one resonance is suitable for low-noise comb generation within the high-power range of the DFB LD because the microresonator FSR is $834\,\text{GHz}$ ($4.79\,\text{nm}$). Fig.\ref{fig:fig2}a shows the optical spectrum of a single soliton generated using the packaged SIL device. The inset in Fig.\ref{fig:fig2}a demonstrates the low-frequency offset noise measured via an electrical spectrum analyzer (ESA, R\&S FPL1014, confirming a low-noise soliton state \cite{Joshi16}. The comb spans a record $600\,\text{nm}$ bandwidth and exhibits a characteristic dispersive wave on the short-wavelength side, located near the zero-dispersion wavelength of $1050\,\text{nm}$ \cite{Okawachi:14,Moille:19}. The operational setpoint was an LD current of $400\,\text{mA}$ (corresponding to $150\,\text{mW}$ of power) and an LD temperature of $35^\circ\text{C}$. The total comb power is 6.4 mW.

\textbf{Bismuth-doped fiber amplifier}

The microcomb has a spectral envelope that concentrates power below 0~dBm level (Fig.\ref{fig:fig2}a), which is below practical per-channel levels \cite{buscaino2021external}.  
We therefore enhance the initial microcomb in the developed wideband bismuth-doped fiber amplifier (BDFA), which is designed as a single-stage, single-pump amplifier. The active medium consists of a newly designed 195 m bismuth-doped fiber (BDF).

The BDF was fabricated by the modified chemical vapor deposition (MCVD) technique as a phosphosilicate silica-core fiber doped with bismuth at a concentration below $0.01$ $mol$\% \cite{nagel1982overview}. The preform was produced using a Heraeus F300 cladding tube, while Si, P, and Bi were deposited from the gas phase. The fabricated fiber exhibited a core--cladding refractive index difference of $2.5\times10^{-3}$, a cutoff wavelength of about $1.1~\mu\mathrm{m}$, and a core diameter of  $5~\mu\mathrm{m}$. Fig.\ref{fig:figSI2} summarizes the characteristics of the developed Bi-doped fiber. Fig.\ref{fig:figSI2}a shows the refractive-index profile, Fig.\ref{fig:figSI2}b presents the measured wavelength-dependent fiber loss, and Fig.\ref{fig:figSI2}c shows the measured \(\mathrm{Bi_2O_3}\) and \(\mathrm{P_2O_5}\) concentration profiles. The refractive-index profile exhibits an M-shaped radial distribution, with a maximum core--cladding index difference of $2.5\times10^{-3}$, corresponding to an estimated \(\mathrm{P_2O_5}\) concentration of \(\sim 3.5\)~mol\%. The Bi concentration is comparatively low, approximately $7\times10^{-3}$~mol\%.

The SIL microcomb and BDFA experimental setup is shown in Fig.\ref{fig:fig4}a. The BDF is counter-pumped at 1150~nm via a WDM coupler/filter that combines the pump and the signal. The 1150~nm pump module is based on 915~nm uncooled multimode semiconductor laser diode operating at 7~A  pump diode current and a high--doped Yb-doped fiber conversion stage. At the amplifier input, an additional WDM filter and isolator were inserted to suppress residual pump and spontaneous emission light and protect upstream components. 

The original microcomb (Fig.\ref{fig:fig2}a)  exhibited a  \(\mathrm{sech}^{2}\)--shaped spectral envelope characteristic of a coherent soliton \cite{Kippenberg2018}, with per-line powers ranging from \(-17~\mathrm{dBm}\) at 1262~nm to \(2~\mathrm{dBm}\) at the 1312~nm pump line. The strongest line at 1312 nm corresponds to the pump, which was not suppressed by any filters. The reduced powers of comb lines \(-2\) (1307~nm) and \(-3\) (1302~nm), approximately \(-11\) and \(-17\)~dBm, respectively, are associated with an avoided mode crossing in the microresonator \cite{Yi2017}.

The original microcomb (Fig.\ref{fig:fig2}a)  exhibited a  \(\mathrm{sech}^{2}\)--shaped spectral envelope characteristic of a coherent soliton \cite{Kippenberg2018}, with per-line powers ranging from \(-17~\mathrm{dBm}\) at 1262~nm to \(2~\mathrm{dBm}\) at the 1312~nm pump line. The strongest line at 1312 nm corresponds to the pump, which was not suppressed by any filters. The reduced powers of comb lines \(-2\) (1307~nm) and \(-3\) (1302~nm), approximately \(-11\) and \(-17\)~dBm, respectively, are associated with an avoided mode crossing in the microresonator \cite{Yi2017}.
The optical spectrum of the BDFA amplified microcomb is presented in Fig.\ref{fig:fig2}b. The red dashed line marks the 0~dBm level and shows that all O--band microcomb lines exceed this threshold after amplification by BDFA with a total output power of 508~mW. 

As shown in Fig.\ref{fig:fig2}b, the BDFA provides 14.3--20~dB of gain for 21 equidistant comb lines in the O-band, while effectively maintaining the input microcomb's spectral envelope, resulting in output comb line powers ranging from 1 to 22~dBm. The amplifier provides approximately 20~dB of gain at the center of the microcomb spectrum. Remarkably, despite a 19~dB (from -17dBm to +2 dBm) variation in the initial power of the microcomb lines, the amplifier maintains a gain flatness of  3~dB over a 70~nm bandwidth (1270--1340~nm), encompassing 15 comb lines. The wavelength-resolved noise figure ranges from 6.5 to 7.9~dB across all measured comb lines (Fig.\ref{fig:fig4} b). The NF was calculated after subtracting the contribution of the input comb source, so that the reported values represent the excess noise introduced by the BDFA rather than the intrinsic RIN or excess noise of the input microcomb.
Fig.\ref{fig:fig4}c shows the comparison of the initial (blue dots) and amplified microcomb (green dots) optical signal-to-noise ratio (OSNR). The non--uniform OSNR across the band is primarily determined by the initial microcomb spectral envelope and is largely preserved after amplification, with additional corrections at the band edges due to BDFA-amplified spontaneous emission. The output OSNR is above 30~dB for all comb lines (0.1~nm RBW), and reaches \(\sim\)50~dB for the central pump line at 1312~nm.

Fig.\ref{fig:fig4}c shows the comparison of the initial (blue dots) and amplified microcomb (green dots) optical signal-to-noise ratio (OSNR). The non--uniform OSNR across the band is primarily determined by the initial microcomb spectral envelope and is largely preserved after amplification, with additional corrections at the band edges due to BDFA-amplified spontaneous emission. The output OSNR is above 30~dB for all comb lines (0.1~nm RBW), and reaches \(\sim\)50~dB for the central pump line at 1312~nm. 

\begin{figure*}[!t]
\centering
\includegraphics[width=1.05\textwidth]{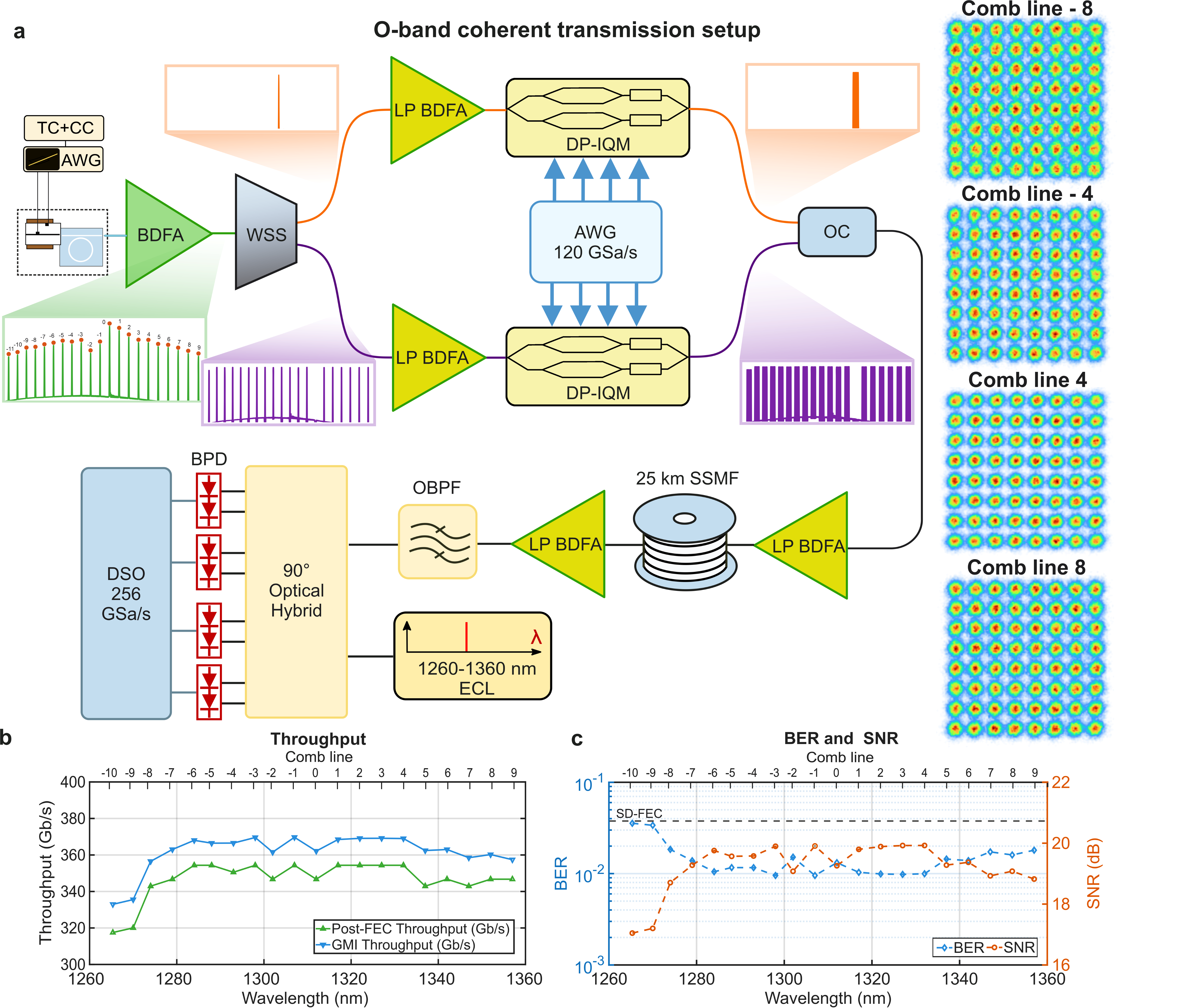}
\caption{\textbf{O-band coherent WDM transmission using the amplified SIL microcomb.}
\textbf{a}, Experimental setup for 25-km coherent transmission. The amplified microcomb is equalised and separated by an O-band wavelength-selective switch (WSS) into a selected channel under test and the remaining comb lines. The two branches are independently amplified and modulated using dual-polarisation IQ modulators (DP-IQMs) driven by a 120~GSa/s arbitrary waveform generator (AWG), recombined using an optical coupler (OC), and transmitted through 25-km of standard single-mode fiber (SSMF). At the receiver, the channel under test is selected by an optical band-pass filter (OBPF) and coherently detected using a tunable external-cavity laser (ECL), a 90$^{\circ}$ optical hybrid and balanced photodetectors (BPDs). Representative recovered DP-64QAM constellations are shown for comb lines $-8$, $-4$, $4$ and $8$.
\textbf{b}, GMI-estimated and post Forward Error Correction (post-FEC) throughput for the characterised comb lines. The post-FEC throughput was obtained using the  DVB-S2 LDPC decoding procedure
described in Methods.
\textbf{c}, Measured Signal-to-Noise Ratio (SNR, orange line, right axis) and pre-FEC Bit Error Rate (BER,  blue line, left axis), as a function of wavelength and comb-line number. The dashed line indicates an SD-FEC pre-FEC BER reference level of
$3.85\times10^{-2}$ and is shown for reference.}

\label{fig:fig5}
\end{figure*}

\textbf{Noise characteristics}

 Fig.\ref{fig:fig4}d presents the frequency noise of the free--run and SIL DFB laser.
The frequency noise spectral density was measured using a self-heterodyne system with a fiber delay line and an AOM modulated at 200~MHz. The intrinsic (Lorentzian) linewidths were extracted directly from the white-noise floor level $S_f$ using the relation $\Delta\nu = \pi S_f$. For the free--running DFB laser, the white-noise floor corresponds to a linewidth of $\sim$ 2~MHz, which is reduced to 2.73~kHz in the SIL regime--representing an intrinsic linewidth suppression by a factor of $\sim$ 735. This narrowing factor is well described by the analytical SIL theory \cite{kondratiev2017self}, being inversely proportional to the microresonator quality factor squared (see Methods). After amplification by the BDFA, the intrinsic Lorentzian linewidth increases moderately to  3.2~kHz, indicating only a limited additional phase-noise contribution from the amplification stage.
Furthermore, the high-$Q$ microresonator allows the SIL DFB laser to reach the thermorefractive noise (TRN) limit at high offset frequencies (above 200 kHz). At lower offset frequencies (below 200 kHz), the frequency noise is primarily dominated by thermal drift and mechanical coupling instability within the package \cite{KONDRATIEV2018,Huang2019}. Notably, the SIL mechanism provides uniform frequency noise reduction across all microcomb lines \cite{xiang2021laser}.

 Fig.\ref{fig:fig4}e shows the relative intensity noise (RIN) of the amplified microcomb lines used as carriers in the subsequent transmission experiments. The RIN spectral density averaged over 1--100~MHz is –135.4~dBc/Hz and –136.2~dBc/Hz for comb lines –6 and 6, respectively. For comb line 1 (1317~nm), which is one of the closest lines to the pump, this value increases to –133.1~dBc/Hz due to a local mode--crossing in the microresonator. The absence of elevated low-frequency RIN indicates weak pump-to-signal intensity-noise transfer in the counter-pumped configuration. These results support the use of a multimode laser diode without active TEC control as the pump source, simplifying amplifier control, and demonstrate the performance of the system for coherent transmission.

 \textbf{Coherent transmission}

To evaluate the amplified SIL microcomb as a multi-wavelength source for coherent optical communications, we performed a 25-km single-span O-band WDM transmission experiment using 32~GBd dual-polarisation 64--quadrature--amplitude modulation (DP-64QAM) \cite{pfeifle2014coherent,geng2022coherent,marzban2026quantum}. The experimental configuration is shown in Fig.~\ref{fig:fig5}a. 
The amplified microcomb was first passed through a developed O-band wavelength-selective switch (WSS), which equalised the comb-line powers, separated one selected comb line as the channel under test (CUT) from the remaining comb lines, and routed the two optical paths independently. The CUT and the remaining comb lines were then modulated using separate dual-polarisation in-phase/quadrature modulators (DP-IQMs), both driven by a 120~GSa/s arbitrary waveform generator with independent data patterns. The selected comb line carried an independently generated 32~GBd DP-64QAM signal, while the remaining available comb lines were simultaneously modulated and used as WDM loading channels. After modulation, the CUT and WDM-loading branches were recombined using an optical coupler (OC). This configuration enables line-resolved assessment of the microcomb carriers under simultaneous WDM loading using a phase-sensitive modulation format \cite{webb1994modern}. The WSS, OC and DP-IQMs introduce wavelength-dependent insertion losses across the O-band. The two branches were therefore amplified independently using low-power BDFAs to compensate for losses introduced by the WSS and subsequent transmitter components~\cite{Donodin2026ECOC,emmerich2026characterization}. After modulation and recombination of the two branches, an additional low-power BDFA was used to compensate for the remaining transmitter losses and establish a controlled launch power into the fiber span.

The use of auxiliary optical amplifiers to compensate for transmitter, propagation and receiver-side losses is routinely employed in coherent transmission testbeds, including long- and short-haul fiber links, optical wireless links and back-to-back configurations \cite{shu2022microcomb,tao2025ultrabroadband,marzban2026quantum}. These auxiliary BDFAs are part of the transmission testbed and are not part of the proposed SIL microcomb source.

At the receiver, the channel under test was selected using a tunable optical band-pass filter (OBPF) and coherently detected using a wavelength-tunable external-cavity laser operating over 1260--1360~nm as the local oscillator. The received signal was detected using a $90^{\circ}$ optical hybrid and balanced photodetectors and digitised by a 256~GSa/s digital storage oscilloscope for offline digital signal processing.
The transmission performance was evaluated successively for 20 microcomb lines spanning approximately 1266--1357~nm. The shortest-wavelength comb line near 1262~nm could not be characterised because it lies outside the operating wavelength range of the WSS used in the transmission testbed. For each wavelength, the selected line was used as the CUT while the remaining 19 available comb lines were transmitted simultaneously as WDM loading channels.

Figure~\ref{fig:fig5}b presents the wavelength-dependent achievable throughput estimated from the generalized mutual information (GMI), together with the corresponding  Post-Forward Error Correction (post-FEC) throughput, the final, usable data after the repairing errors. The GMI-estimated throughput ranges from approximately 333 to 370~Gb/s per comb line, while the post-FEC throughput ranges from approximately 318 to 355~Gb/s per line. Summed over the 20 experimentally characterised comb lines, the aggregate throughput is $\sim$7.23~Tb/s based on GMI and $\sim$6.94~Tb/s after FEC decoding. The aggregate post-FEC throughput is approximately 4.0\% below the GMI-estimated throughput.

The wavelength dependence of the transmission performance is further illustrated by the measured pre-FEC BER (before any error correction takes place) and SNR in Fig.~\ref{fig:fig5}c.The dashed line indicates the pre-FEC BER threshold of
$3.85\times10^{-2}$ for a 25.5\%-overhead soft-decision FEC (SD-FEC)
scheme and is shown for reference only \cite{Bernal2024NatCommun}. The post-FEC throughput in Fig.~\ref{fig:fig5}~b was obtained using the DVB-S2 LDPC decoding procedure described in the Methods section. The lowest performance is observed at the short-wavelength edge of the measured band, where the reduced SNR is accompanied by an increased BER and lower achievable throughput. From approximately 1270~nm towards the center and long-wavelength side of the O-band, the SNR remains close to 19--20~dB and the post-FEC throughput is typically above 340~Gb/s per line. 

Representative recovered 64-QAM constellations for comb lines $-8$, $-4$, $4$ and $8$ are shown in Fig.~\ref{fig:fig5}a. Well-resolved constellation points are observed across the full O-band, confirming successful coherent data transmission over ~ 100 nm spectral range. Higher aggregate capacities could in future be obtained using higher symbol rates, more spectrally efficient modulation, or smaller-FSR microresonators where denser WDM operation is required. These results provide system-level validation of the amplified SIL microcomb as a broadband coherent carrier source for O-band optical communication systems. 

\section{Conclusion and Outlook}\label{sec3}
We have demonstrated a high-power O-band microcomb platform that combines self-injection-locked broadband microcombs in a Si\textsubscript{3}N\textsubscript{4} microresonator with wideband, single-stage bismuth-doped fiber amplifier. Operating with an 834~GHz free spectral range, the SIL microcomb spans 1050-1650~nm. Using the designed bismuth-doped phosphosilicate fiber, we simultaneously amplified 21 O-band comb lines to powers exceeding 0~dBm per line across the full 1260--1360~nm transmission window, with 5~dB amplifier gain variation across the $\sim$100-nm O-band and approximately 3~dB gain variation over 70~nm (1270--1340~nm), without the use of gain-flattening filters in the BDFA. The amplified low-noise microcomb source maintains an OSNR above 30~dB across the O-band, low RIN and an intrinsic linewidth of 3.2~kHz after amplification.
We further validated the amplified microcomb in a 25-km coherent WDM transmission experiment using 32~GBd dual-polarisation 64-QAM. The amplified comb lines, spanning within the operating range of the WSS, were experimentally characterised while the remaining available comb lines were simultaneously transmitted as WDM loading channels. The measured aggregate GMI-estimated throughput reached $\sim$ 7.23~Tb/s for twenty comb lines, with a post-FEC throughput of $\sim$ 6.94~Tb/s, corresponding to an implementation penalty of $\sim$ 4\%. Well-resolved constellations and consistent transmission performance across widely separated wavelengths confirm the suitability of the amplified SIL microcomb as a broadband coherent O-band carrier source.

Microcomb-based WDM transmitters provide multiple wavelength carriers from a single optical source, offering a route to reduced source count and simplified wavelength management compared with arrays of independent lasers. The demonstrated combination of broadband SIL microcomb generation, O-band fiber amplification and multi-terabit coherent transmission provides a platform for scalable short-reach optical interconnects and other photonic systems requiring broadband multi-wavelength operation.
\section*{Methods}

\textbf{Self--injection locking linewidth narrowing}

The theory of linear SIL and representative early experiments are described in Refs \cite{ oraevsky2001frequency,kondratiev2017self,kondratiev2020numerical}, while nonlinear SIL dynamics with comb generation are discussed in Ref. \cite{voloshin2021dynamics}. The SIL linewidth narrowing can be described as  $\delta\omega \approx \delta\omega_{\rm freerun} (Q_{\rm LD} / Q_{\rm MR})^2 \cdot [16r^2(1+\alpha_g^2)]^{-1}$,
where $\delta\omega_{\rm freerun} / 2 \pi$ is the linewidth of the free running laser, $Q_{MR}$ is the microresonator quality factor, $\omega / 2 \pi$ is the light frequency. The parameter $\alpha_g$ is the linewidth enhancement factor, given by the ratio of the variation of the real refractive index to the imaginary refractive index of the laser diode active region in response to a carrier density fluctuation\cite{kondratiev2017self}.

\textbf{Amplified Si\textsubscript{3}N\textsubscript{4} microcomb characterization}
The noise figure of BDFA was obtained using Source Subtraction Technique which applicable for multichannel characterization of fiber amplifiers \cite{baney2000theory}:

\begin{equation}
NF(\lambda)
=
10\log_{10}\!\left[
\frac{2\,\rho_{\mathrm{ASE}}(\lambda)}{G(\lambda)\,h\nu}
+
\frac{1}{G(\lambda)}
\right],
\end{equation}

where \(G(\lambda)\) is the (linear) gain, \(\nu=c/\lambda\) is the optical frequency, and \(\rho_{\mathrm{ASE}}(\lambda)\) is the ASE spectral density at the amplifier output. 
The OSNR was evaluated directly from the output spectrum using a real optical spectrum analyzer (OSA Yokogawa AQ6370D) resolution bandwidth that varies from  0.126 to 0.109 nm across the O-band, and consistent averaging settings across all wavelengths. 

For the RIN measurements, individual filtered lines from the amplified microcomb were coupled into an amplified InGaAs photoreceiver (Newport 1611FC-AC). The receiver inherently separates the signal into DC and AC components via dedicated outputs. The DC monitor port was continuously tracked with a digital multimeter to ensure a strictly constant average incident optical power across all measurements. Simultaneously, the RF noise power was acquired from the AC-coupled output using an electrical spectrum analyzer (Rohde and Schwarz FPL1014).

Frequency-noise measurements were performed using the in-phase and quadrature (I/Q) data exported from the electrical spectrum analyzer. The complex signal was reconstructed as $z(t)=I(t)+jQ(t)$, and the instantaneous phase was obtained from the unwrapped argument of this signal. A linear trend was removed from the phase trace to eliminate the average carrier offset and retain only the phase fluctuations. The phase-noise power spectral density, $S_{\phi}(f)$, was then calculated from the residual phase using Fourier analysis with a Hann window and, where required, Welch averaging. For measurements performed with a delayed interferometric discriminator, the obtained spectrum was corrected by the interferometer transfer function, $[2\sin(\pi f\tau)]^{2}$, where $\tau$ is the interferometer delay. The frequency-noise power spectral density was finally obtained as $S_{\nu}(f)=f^{2}S_{\phi}(f)$ \cite{schiemangk2014accurate}. 


\textbf{Transmission performance estimation}

The transmission performance was evaluated using both the generalized mutual information (GMI) and post-FEC decoding. The GMI was calculated from the bit-wise log-likelihood ratios (LLRs) of the recovered DP-64QAM symbols and summed over the two polarisations. The corresponding GMI-estimated throughput was calculated as

\begin{equation}
R_{\mathrm{GMI}} = R_{\mathrm{s}} \times \mathrm{GMI},
\end{equation}

where $R_{\mathrm{s}}$ is the symbol rate and the GMI is expressed in bits per dual-polarisation symbol. For the 32~GBd DP-64QAM signals used here, the maximum uncoded information rate is therefore
$32\times 12=384$~Gb/s per comb line.
For the post-FEC throughput, we adopted the decoding approach used in Ref.~\cite{puttnam2024402}, employing low-density parity-check (LDPC) codes from the DVB-S2 standard \cite{second2013digital} with puncturing to provide finer code-rate granularity. For each measured comb line, progressively higher code rates were evaluated, and the highest code rate yielding a post-decoding BER below $10^{-5}$ was selected. A further 1\% overhead for an outer hard-decision FEC code was assumed to remove the residual errors \cite{millar2016design}. The resulting net post-FEC throughput was calculated as

\begin{equation}
R_{\mathrm{FEC}}
=
R_{\mathrm{s}}
\times 2\log_{2}(M)
\times R_{\mathrm{c}}
\times 0.99,
\end{equation}

where $M=64$ is the modulation order and $R_{\mathrm{c}}$ is the highest successfully decoded DVB-S2 code rate. The difference between the GMI-estimated and post-FEC throughputs therefore represents the implementation and coding penalty of the practical FEC scheme.

\section*{Acknowledgements}

For the purposes of open access the authors have applied a Creative Commons Attribution (CC BY) licence to any Author Accepted Manuscript (AAM) version arising from this submission.

This work was supported by the UK Engineering and Physical Sciences Research Council (EPSRC) grant number EP/W002868/1 and EP/X031918/1 (IMPAC), A.D. was supported by the RAEng under Research Fellowship, parts of the work are supported by the German Federal Ministry of Research, Technology and Space (BMFTR) under the HYPERCORE project(grant ID 16KIS2097K). 
\section*{Author contributions}
D.S., N.G.P., and S.K. performed the microcomb characterization. V.M., J.L., and D.J.D. designed and characterized the Bi-doped fiber. D.S., A.D., V.M., J.L., and D.J.D. contributed to the development of the BDFA. D.S., A.D., D.J.E., R.E., R.S.L., H.F. C.S., R.F., Y.W., and T.T. designed the coherent transmission setup and contributed to the experiments.  D.S., N.G.P., A.D., and D.J.E. analyzed the data. N.G.P., M.K., J.D.J., and S.K.T. supervised the project. D.S., N.G.P., A.D., D.J.E., V.M., and S.K.T. wrote the manuscript with input from all authors.

\section*{Competing Interests}
We declare that none of the authors have competing  
interests but disclose for transparency that N.G.P., J.D.J. and M.K. are cofounders of Enlightra.

\noindent $^\dagger$ These authors contributed equally to this work.

\bibliography{references.bib}

\end{document}